\documentclass[aps,prb,twocolumn,showpacs,superscriptaddress,amsmath,amssymb,
tightenlines,floatfix]{revtex4}

\usepackage[a4paper,left=1.5cm,right=1cm,top=3cm,bottom=2cm]{geometry}
\usepackage{graphicx}
\usepackage{dsfont}
\usepackage[usenames,dvipsnames]{xcolor}
\usepackage{amsbsy}

\newcommand {\Sec}[1] {Section~\ref{#1}}
\newcommand {\Eq}[1] {Eq.~\ref{#1}}
\newcommand {\App}[1] {Appendix~\ref{#1}}

\newcommand {\Fig}[1] {Figure~\ref{#1}}

\newcommand{\beq}{\begin{equation}}
\newcommand{\eeq}{\end{equation}}

\newcommand{\cthirteen}{$^{13}$C}

\newcommand{\sitwonine}{$^{29}$Si}
\newcommand{\sitwoeight}{$^{28}$Si}

\newcommand{\beqa}{\begin{eqnarray}}
\newcommand{\eeqa}{\end{eqnarray}}

\newcommand{\ket}[1]{\left| #1 \right\rangle}
\newcommand{\bra}[1]{\left\langle #1 \right|}

\newcommand{\ttwo}{$T_2$}

\renewcommand{\Eq}[1]{Eq.~(\ref{#1})}
\renewcommand{\Fig}[1]{Fig.~\ref{#1}}
\newcommand{\Bfig}[1]{Figure~\ref{#1}}
\newcommand{\Beq}[1]{Equation~(\ref{#1})}
\newcommand{\Table}[1]{Table.~(\ref{#1})}

\renewcommand{\ket}[1]{\left|{#1}\right\rangle}
\newcommand{\sket}[1]{|{#1}\rangle}
\renewcommand{\bra}[1]{\left\langle{#1}\right|}

\newcommand{\avg}[1]{\left\langle{#1}\right\rangle}

\newcommand{\backtoback}[2]{\left\langle{#1}|{#2}\right\rangle}

\begin{document}

\title{Quantum bath-driven decoherence of mixed spin systems}

\author{S.~J.~Balian}
\affiliation{Department of Physics and Astronomy, University College London,
Gower Street, London WC1E 6BT, United Kingdom}

\author{Gary~Wolfowicz}
\affiliation{London Centre for Nanotechnology, University College London,
London WC1H 0AH, UK}
\affiliation{Department of Materials, Oxford University, Oxford OX1 3PH, United Kingdom}

\author{John~J.~L.~Morton}
\affiliation{London Centre for Nanotechnology, University College London,
London WC1H 0AH, UK}
\affiliation{Department of Electronic \& Electrical Engineering, University
College London, London WC1E 7JE, UK} 

\author{T.~S.~Monteiro}
\email{t.monteiro@ucl.ac.uk}
\affiliation{Department of Physics and Astronomy, University College London, 
Gower Street, London WC1E 6BT, United Kingdom}

\date{\today}

\begin{abstract}

The decoherence of mixed electron-nuclear spin qubits is a topic of great current importance, but understanding is still lacking: while important decoherence mechanisms for spin qubits arise from quantum spin bath environments with slow decay of correlations, the only analytical framework for explaining observed sharp variations of decoherence times with magnetic field is based on the suppression of classical noise. Here we obtain a general expression for decoherence times of the central spin system which exposes significant differences between quantum-bath decoherence and decoherence by classical field noise. We perform measurements of decoherence times of bismuth donors in natural silicon using both electron spin resonance (ESR) and nuclear magnetic resonance (NMR) transitions, and in both cases find excellent agreement with our theory across a wide parameter range. The universality of our expression is also tested by quantitative comparisons with previous measurements of decoherence around `optimal working points' or `clock transitions' where decoherence is strongly suppressed. We further validate our results by comparison to cluster expansion simulations.

\end{abstract}



\maketitle

\section{Introduction}

Understanding decoherence of electron spins in the solid state is both of practical importance, to exploit them in quantum technologies such as quantum computers,\cite{DeSousa2003} and also of fundamental interest in addressing questions such as how decoherence by quantum baths (associated with back-action and environment-memory effects)\cite{Breuer2002,Maniscalco2006,Mazzola2012} differs from classical noise sources.

One of the leading sources of electron spin decoherence is due to coupling to other spins in the environment. In some cases, the host material is highly rich in nuclear spins (such as III-V semiconductors) limiting electron spin decoherence times (\ttwo) to less than a microsecond,\cite{Koppens2008} while in other cases a small natural abundance of nuclear spins (such as 5$\%$ of \sitwonine\ in silicon, or 1$\%$ \cthirteen\ in diamond) limits \ttwo\ to a few hundred microseconds.\cite{Tyryshkin2003, George2010, Gaebel2006} Even when the nuclear spins have been almost completely removed (such as in enriched \sitwoeight), \ttwo\ is then typically limited by coupling to other electron spins in the environment.\cite{Tyryshkin2012} The decoherence dynamics of spins interacting with a quantum bath of other spins is therefore of much interest.\cite{Takahashi2008,DeLange2012,Zhao2011,Friedemann2012}

More recently, systems with substantial electron-nuclear spin mixing have been attracting considerable attention, especially due to the presence of `clock transitions' or `optimal working points' (OWPs) where the qubit shows enhanced robustness to decoherence\cite{Mohammady2010,Mohammady2012,Balian2012,Wolfowicz2013} and \ttwo\ varies by orders of magnitude. A large number of important defects in the solid state possess such mixing, including donors in silicon,\cite{Steger2012,Morley2013,Wolfowicz2012} NV centres in diamond,\cite{Zhao2012} transition metals in II-VI materials\cite{George2013} and rare-earth dopants in silicates.\cite{Fraval2005} Earlier studies of other systems which have OWPs but are primarily affected by classical noise,\cite{Vion2002} led to theoretical analyses of the dependence of $T_2$ on field noise,\cite{Ithier2005,Martinis2003} both at and far from OWPs. In contrast, no comparable general analytical expressions for \ttwo\ have yet been obtained for spin systems decohered by quantum baths.

In this paper, we examine decoherence of donors in silicon caused by nuclear spin diffusion. We show that the spin dynamics separate naturally into terms acting on very different timescales, allowing us to obtain an analytical form for \ttwo\ which i) exposes important and qualitative differences between the quantum bath-driven and typical classical noise-driven decoherence and ii) is fully generalizable to mixed electron-nuclear spin systems. The decoherence time is given as a function of key mixing terms:
$T_2 \simeq \overline{C}(\theta) \left(|P_u| + |P_l|\right){\left|P_u-P_l\right|}^{-1}$
where the significant parameter here is $P_i \equiv \bra{i} \hat{S}^z \ket{i}$, corresponding to the electron $S^z$ component of the upper ($\ket{i=u}$) and lower ($\ket{i=l}$) eigenstates for the transition $\ket{u} \to \ket{l}$, noting $P_i$ is a simple analytical function of magnetic field $B$. The constant, $\overline{C}(\theta)$, depends only on magnetic field orientation, the density of nuclear spin
impurities and their gyromagnetic ratio. The expression is shown to give excellent agreement with numerics and experimental data for both ESR and NMR-type transitions as well as OWP regimes.

The paper is organized as follows. In \Sec{Sec:Decoherence}, we describe the central spin decoherence problem of spin diffusion and briefly review established numerical methods for obtaining $T_2$.
In \Sec{Sec:derivation}, we present the derivation of our $T_2$ expression. The importance of separation of timescales is explained
and by employing a strong coupling approximation a closed-form $T_2$ formula is obtained valid for both mixed and unmixed spins.
In \Sec{Sec:Bismuth} we test the formula against numerics, new ESR and NMR data as well as previously obtained OWP data; we show that it yields excellent agreement throughout. Finally, our findings are summarized in \Sec{Sec:Conclusion}.

\section{Central spin decoherence}\label{Sec:Decoherence}

\subsection{Spin Hamiltonian}

We begin with the Hamiltonian for the central spin decoherence problem:
\begin{equation}
 \hat{H}_{\text{tot}} =
 \hat{H}_{\text{CS}}
 + \hat{H}_{\text{int}}
 + \hat{H}_{\text{bath}},
\label{Eq:Htotal}
\end{equation}
where $\hat{H}_{\text{CS}}$ is the central spin system Hamiltonian including all internal nuclear and electronic degrees of freedom, while $\hat{H}_{\text{bath}}$ is the bath Hamiltonian and $\hat{H}_{\text{int}}$ describes the interaction of the central spin with the bath.

We consider the situation where the central spin interacts with a nuclear bath (e.g.\ spin-$1/2$ $^{29}$Si impurities) through the contact hyperfine interaction
\begin{equation}
\hat{H}_{\text{int}} = \sum_a \hat{{\bf S}} {\bf J}_a  \hat{{\bf I}}_{a},
\label{Eq:Hint}
\end{equation}
where $\hat{{\bf S}}$ represents the central electron spin, ${\bf J}_a$ is the contact hyperfine tensor and $a$ labels the bath spins $\hat{{\bf I}}_{a}$.
One can also consider other types of interaction, where $\hat{H}_{\text{int}}$ includes both electronic as well as nuclear ($\hat{{\bf I}}$) terms of the central spin.

Finally, the bath Hamiltonian consists of nuclear Zeeman terms and dipolar coupling among bath spins:
\begin{eqnarray}
\hat{H}_{\text{bath}} &=& \hat{H}_\text{D} + \hat{H}_\text{NZ}, \nonumber\\
\hat{H}_\text{NZ}     &=& \sum_{a} \gamma_N B \hat{I}_a^z, \nonumber\\
\hat{H}_\text{D}      &=& \sum_{a < b} \hat{{\bf I}}_{a} {\bf D}({\bf r}_{ab}) \hat{{\bf I}}_{b},
\label{Eq:Hbath}
\end{eqnarray}
where $\gamma_N$ is the nuclear (bath) gyromagnetic ratio and ${\bf r}_{ab}$ denotes the relative position of bath spins at lattice sites $a$ and $b$.
The components of the dipolar tensor ${\bf D\left({\bf r}\right)}$ are given by
\begin{equation}
D_{i j}({\bf r}) = \frac{\mu_0\hbar\gamma_N^2}{4\pi r^3}\left(\delta_{i j} - \frac{3 r_{i} r_{j}}{r^2}\right),
\label{Eq:Dipolar}
\end{equation}
where $\delta_{i j}$ denotes the Kronecker delta, $\mu_0 = 4\pi \times 10^{-7}$~NA$^{-2}$ and
$i,j = \{x,y,z\}$.

\begin{figure}[t]
\includegraphics[width=3.5in]{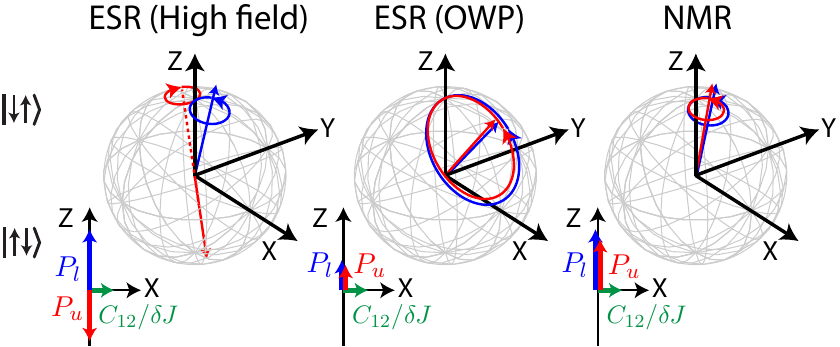}
\caption{(color online) Illustration of the evolution of the bath states in the Hilbert space spanned by $\{\ket{\uparrow \downarrow},\ket{\downarrow \uparrow}\}$ under the influence of their dipole coupling ($C_{12}$) and their mutual detuning caused by interaction with the central spin (see text for full details). At both OWPs and NMR-type transitions, bath trajectories correlated with the upper and lower central spin states follow similar trajectories and hence decoherence is suppressed compared to ESR-type transitions. However, at ESR OWPs, $|P_{u,l}| \simeq 0.05$ leads to a larger trajectory and proportionately shorter \ttwo\ values relative to NMR-type transitions.
}
\label{Fig:Pseudospins}
\end{figure}

\subsection{Coherence decay for pure dephasing}

The decay in coherence of the central spin can be related to its entanglement with the bath.
A measurement of \ttwo\ begins by applying a $\pi/2$ pulse to produce the initial state
$\tfrac{1}{\sqrt{2}}\left(\ket{u} + \ket{l}\right)
\otimes \ket{\mathcal{B}(0)}$,
where $\ket{u}$ and $\ket{l}$ are a pair of eigenstates of $\hat{H}_{\text{CS}}$ and $\ket{\mathcal{B}(0)}$ is
the initial state of the bath.
After a finite time delay $t$, the state evolves into the entangled state
\begin{equation}
\ket{t} = \tfrac{1}{\sqrt{2}} \left(e^{-i E_u t}\ket{u} \otimes \ket{\mathcal{B}_u\left(t\right)} + e^{-i E_l t}\ket{l} \otimes \ket{\mathcal{B}_l\left(t\right)}\right),
\label{Eq:TimeEvolution}
\end{equation}
where $E_u$ and $E_l$ are the energies associated with $\ket{u}$ and $\ket{l}$ respectively.
We consider the situation of pure dephasing, i.e.\ we assume that the effect of $\hat{H}_{\text{int}}$ on the central spin states remains negligible during evolution under $\hat{H}_\text{tot}$.
The complex off-diagonal of the central spin density matrix $\mathcal{L}^{u\to l}(\ket{\mathcal{B}(0)},t)$ is proportional to $\avg{\hat{\sigma}^x} \pm i\avg{\hat{\sigma}^y}$ where the $\hat{\sigma}^x$ and $\hat{\sigma}^y$ are Pauli operators in the $\{\ket{u},\ket{l}\}$ basis. Hence, $|\mathcal{L}^{u\to l}(\ket{\mathcal{B}(0)},t)|$ is proportional to the signal in an experiment probing the transverse magnetization (e.g.\ free induction decay (FID) or Hahn spin echo).
For pure dephasing, this coherence decay is simply given by
$|\mathcal{L}^{u\to l}(\ket{\mathcal{B}(0)},t)| = |\backtoback{\mathcal{B}_l(t) }{ \mathcal{B}_u(t)}|$.
A measurement probing $\avg{\hat{\sigma}^x} \pm i\avg{\hat{\sigma}^y}$ will experience decay if
$|\mathcal{L}^{u\to l}(\ket{\mathcal{B}(0)},t)| \neq 1$. In effect, obtaining the decoherence rates involves calculation of the time-dependent overlap between bath states correlated with the upper and the lower central spin states.

\subsection{Numerical simulation of $T_2$}

Recent advances in solving central spin decoherence problems,\cite{Yao2006,Yao2007,Maze2008,Witzel2010,Witzel2012} including the cluster correlation expansion (CCE),\cite{Yang2008_2008E_2009} have enabled realistic numerical simulations of the joint system-bath dynamics. 
In the CCE and analogous formalisms, $\hat{H}_{\text{tot}}$ is diagonalized for sets or ``clusters'' of bath spins of varying sizes. The coherence decay is obtained from a product over all cluster contributions in the bath.

In previous studies of nuclear spin diffusion of donor spins in silicon,
pair correlations (pairs of spin clusters or `2-clusters'),\cite{Yao2006} were found to dominate
the decoherence, with clusters of three or more bath spins making only a minor
contribution.\cite{Abe2010,Witzel2010} In this case, the expansion is simply a product over all pairs,
\begin{equation}
\mathcal{L}^{u \to l}(t) = \prod_{n} \mathcal{L}^{u \to l}_{n}(t),
\end{equation}
where $\mathcal{L}_{n}^{u \to l}(t)$ is the complex off-diagonal calculated for the $n$-th spin pair.
These are referred to below as spin pair-correlation decays.

\subsection{Pseudospin model for bath dynamics}

For a variety of spin problems including quantum dots and NV centres (for both one-spin and spin pair clusters),
the bath dynamics for the $n$-th cluster can be treated as precession of either a spin or pseudospin 
about an effective and central spin state-dependent magnetic field (\Fig{Fig:Pseudospins}).\cite{Yang2008_2008E_2009,Yao2006,Yao2007}  
Similarly, for the donors spin qubits in silicon, one may assume $H_{\text{CS}} \gg H_{\text{bath}}$ and thus ignore non-secular terms in $\hat{H}_{\text{int}}$;
the interaction Hamiltonian (\Eq{Eq:Hint}) for the $n$-th cluster reduces to Ising form:
~$\hat{H}^{(n)}_{\text{int}} = \sum_{a=1,2}J^{(n)}_a \hat{S}^z \hat{I}^z_i$, with hyperfine coupling strengths $J^{(n)}_a$.
Considering a spin-$1/2$ bath (with eigenstates $\ket{\uparrow}$ and $\ket{\downarrow}$) and keeping only spin conserving terms, the dipolar interaction given in Equations~(\ref{Eq:Hbath}) and (\ref{Eq:Dipolar})
simplifies to
$\hat{H}^{(n)}_{\text{bath}} = 2 C^{(n)}_{12} \hat{I}_1^z \hat{I}_2^z - \tfrac{C^{(n)}_{12}}{2} (\hat{I}_1^+ \hat{I}_2^- + \hat{I}_1^- \hat{I}_2^+)$, where $C^{(n)}_{12}$ is the dipolar coupling strength between the two bath spins.
Zeeman terms are also excluded from $\hat{H}^{(n)}_{\text{bath}}$
as these do not contribute to decoherence.
Neglecting the effect of $\hat{H}_{\text{int}}$ on the mixing of the central spin states,
the dynamics is governed by $\hat{h}^{(n)}_{i}$ (conditional on the state of the central spin):
\begin{equation}
\hat{h}^{(n)}_{i} \equiv
\bra{i}(\hat{H}_\text{int}^{(n)} + \hat{H}_\text{bath}^{(n)})\ket{i} = -\tfrac{C_{12}}{2} \hat{\mathds{1}} -\tfrac{1}{2}\hat{\boldsymbol\sigma}\cdot {\bf H}^{(n)}_{i},
\label{Eq:Hpseudospin}
\end{equation}
where the effective field is ${\bf H}^{(n)}_i=[C^{(n)}_{12}, 0, P_i\delta_J^{(n)}]$.
Here, $\delta_J^{(n)}\equiv (J^{(n)}_1 - J^{(n)}_2)$ is the difference in hyperfine couplings to the bath while
$\hat{\boldsymbol\sigma}$ is the vector of Pauli matrices in the bath basis $\{\ket{\downarrow \uparrow},\ket{\uparrow \downarrow}\}$.
The identity term is dynamically uninteresting; the dynamics can in fact be
considered simply as a precession about ${\bf H}_i^{(n)}$.
The pseudospin precession rate is $\omega_i^{(n)}=\frac{1}{2}\sqrt{(C_{12}^{(n)})^2+ (P_i\delta_J^{(n)})^2}$, while the angle of ${\bf H}_i^{(n)}$ from the $z$-axis is $\theta_{i}^{(n)} = \tan^{-1}{[C_{12}^{(n)}/(P_i\delta_J^{(n)})]}$.\\

For the mixed electron-nuclear spin systems investigated here, the pseudospin dynamics is
in most respects, quite similar to those investigated previously for electron (unmixed)
 qubits.\cite{Yang2008_2008E_2009,Yao2006,Yao2007,Zhao2012}
However, the main  difference is that in \Eq{Eq:Hpseudospin}, we have 
replaced $\hat{S}^z$ in $\hat{H}_\text{int}^{(n)}$ by the mixing term
$P_i \equiv \bra{i} \hat{S}^z \ket{i}$.
While for an electron, $P_i=\pm 1/2$ is a constant, 
for mixed systems the $P_i(B)$ are strongly field-dependent. We also assume that $\hat{H}_\text{int}^{(n)}$ has negligible effect on the mixing of the central spin states themselves, i.e.\ on $P_i$, since 
$H_\text{CS} \gg H_\text{int}$.
This assumption is reasonable except extremely close to OWPs, where $T_2$ becomes extremely
 sensitive to small fluctuations in $P_i$.

The $n$-th cluster decay for a single spin pair
has been investigated analytically for both the FID and
Hahn echo case.\cite{Yao2006,Zhao2012,Witzel2005}
We emphasize that this is a `one-central spin' FID (without inhomogeneous broadening).
In experiment, \ttwo\ is normally measured using a Hahn echo pulse sequence ($\pi/2 - \pi - \text{echo}$), in order to remove strong enhancements in decoherence arising from static inhomogeneities.
 Although the Hahn echo can suppress some effects of the dynamics,
the `one-central spin' FID and Hahn $T_2$ times are of the same order, differing by at most 
a factor of $\approx 2$, so  we focus our analysis on
 the simpler FID expressions.

\section{Derivation of $T_2$ expression}\label{Sec:derivation}

Although analytical forms for the decays $\mathcal{L}_{n}^{u \to l}(t)$ from spin pairs are known,\cite{Yao2007,Zhao2012}
a closed form for $T_2$, sufficiently accurate for experimental analysis is more difficult.
Each $\mathcal{L}^{u \to l}_{n}(t)$ is an oscillatory
function, with frequencies given in terms of $\omega_u^{(n)}$ and $\omega_l^{(n)}$
and the full decays combines hundreds or thousands of spin pair contributions.

A usual approach is to expand the decay as a power series $|\mathcal{L}_{n}^{u \to l}(t)|= 1- \sum_{p=1} a^{(n)}_{2p} t^{2p}$ and to infer the order of magnitude of $T_2$ from the early time behavior.
 However, for important cases like spin diffusion, $a_{2}^{(n)}=0$ while $a^{(n)}_4 \neq 0$, predicting a $\exp[-a^{(n)}_4 t^4]$ decay,\cite{Witzel2005,Yao2006,Yao2007} 
in contrast to the observed decays of $ \sim \exp[- t^2/T^2_2]$ for typical spin
systems. Thus it appears that in that case, one cannot infer the character of the decay
on timescales $t \sim T_2$ from the short time behavior (i.e. on timescales $t \sim \omega_i^{-1}$ ). 
 One of our key findings is that $T_2$ times sufficiently reliable for
experimental analysis are obtainable analytically if we consider separately, the different frequency
 terms involved in the decays. This is especially important when these terms act on
very different timescales.

\subsection{Separation of timescales}

The evolution of the bath during the free induction decay (FID) of the central spin follows
$\boldsymbol{\mathcal{B}}_i(t)={\bf R}_y(\theta_{i}){\bf R}_z(2\omega_i t) {\bf R}_y^\intercal(\theta_i)\boldsymbol{\mathcal{B}}(0)$
in the matrix representation, where ${\bf R}_y$ and ${\bf R}_z$ represent the usual rotation matrices\cite{Nielsen2010} and $\boldsymbol{\mathcal{B}}(0)$ is the initial bath state in the basis $\{ {(0 \ 1)}^\intercal : \ket{ \uparrow \downarrow} , {(1 \ 0)}^\intercal : \ket{ \downarrow \uparrow} \}$ and in general can be a superposition of $\ket{\uparrow \downarrow}$ and $\ket{\downarrow \uparrow}$. We have dropped the 2-cluster label $n$ for clarity.

The bath overlap for FID follows
\begin{eqnarray}
&&\mathcal{L}_{\text{FID}}^{u \to l}(t) =
\boldsymbol{\mathcal{B}}^\intercal(0)
{\bf T}^*_{ul}(\omega^-, \omega^+,t)
\boldsymbol{\mathcal{B}}(0); \nonumber\\
&&{\bf T}^*_{ul}(\omega^-, \omega^+,t) = \label{Eq:FID} \\
&&{\bf R}_y(\theta_{u})\left( \begin{array}{cc}
e^{i \omega^- t} \cos{\theta^-} & e^{i \omega^+ t} \sin{\theta^-}\\
-e^{-i \omega^+ t}\sin{\theta^-}  & e^{-i \omega^- t}\cos{\theta^-} \end{array} \right) {\bf R}_y^\intercal(\theta_l)\nonumber
\end{eqnarray}
where $\theta^\pm= \frac{1}{2} (\theta_{u}\pm \theta_{l})$ and
$\omega^{\pm}=\omega_u \pm \omega_l$.
We can show that
$\mathcal{L}^{u \to l}_{\text{Hahn}}(2t)=
\boldsymbol{\mathcal{B}}(0)^\intercal
{\bf T}^*_{ul}(\omega^+, \omega^-,t){\bf T}_{ul}(\omega^+, \omega^-,t)\boldsymbol{\mathcal{B}}(0)$,
noting the exchange in order of $\omega^\pm$ relative to the FID
case.
For both FID and Hahn echo, we see that expressions for the decays arise naturally in terms of $\omega^{\pm}$ rather than $\omega_u$ and  $\omega_l$ as is usual.

For example,  for $\boldsymbol{\mathcal{B}}(0)^\intercal = {(0 \ 1)}$ or ${(1 \ 0)}$, the time decay
 for FID ($|\mathcal{L}^{u\to l}_{\text{FID}}(t)|= |\{{\bf T}^*_{ul}(\omega^-, \omega^+,t)\}_{11}|$) is given by
\begin{eqnarray}
|\mathcal{L}^{u\to l}_{\text{FID}}(t)| &=&
\big|D^+e^{-i\omega^- t} + D^-e^{+i\omega^- t} \nonumber \\
           && ~~+ R^+e^{-i\omega^+t} + R^-e^{+i\omega^+t}\big|,        
\label{Eq:FIDupdown}
\end{eqnarray}
where $R^{\pm} = \frac{1}{2}\sin \theta^-(\sin \theta^- \mp \sin \theta^+)$ while
$D^{\pm} =  \frac{1}{2}\cos \theta^-(\cos \theta^- \pm  \cos \theta^+)$.
We then take 
$\left \langle \mathcal{L}^{u \to l}_{\text{FID}}(t)\right\rangle \approx \tfrac{1}{2} + \tfrac{1}{2} |\mathcal{L}^{u \to l}_{\text{FID}}(\ket{\uparrow \downarrow},t)|$
to allow for the fact that approximately half the bath spins are in $\ket{\uparrow \uparrow}$ and $\ket{\downarrow \downarrow}$ states which cannot flip-flop.

We consider  \Eq{Eq:FIDupdown} in three principal limits: (i) for an ESR transition in the high-field regime in which the states are not mixed, $P_u \simeq -P_l$; (ii) for an NMR transition in the high-field regime, or for any transition near an OWP, $P_u \simeq P_l$; and (iii) for an intermediate regime corresponding to a Landau-Zener crossing,\cite{Mohammady2010} where one of the $P_i\simeq0$.

For either (i) or (ii) (\Fig{Fig:Pseudospins}), since $|P_u| \simeq |P_l|$ then $\omega_u \simeq\omega_l$ and thus $\omega^+/\omega^- \gg 1$. For (i), for timescales $\ll (\omega^+)^{-1}$, we neglect the slow oscillations (i.e.\ those in $\omega^-$) in \Eq{Eq:FIDupdown}, which contribute only on very long timescales. Then, at short times and expanding the decay as a power series to leading order,
$|\mathcal{L}_{\text{FID}}^{u \to l}(t)| \approx 1-(t/T_2^{(n)})^2 \simeq \exp{[-(t/T_2^{(n)})^2]}$,
we obtain the $n$-th cluster contribution from only these fast terms:
\begin{equation}
\frac{1}{T_2^{(n)}}  \simeq \frac{1}{2}\left|\sin \theta_u -\sin \theta_l\right| \frac{\omega^+}{2},
\label{Eq:FIDWeights}
\end{equation}
noting that the first term is the difference in precession radii of the pseudospins, while the second term denotes the average precession rate.
In terms of the usual flip-flop models, we note that a larger precession radius corresponds to a larger flip-flop amplitude, while a larger precession frequency
corresponds to a higher flip-flop frequency.

For (ii), $\omega^+/\omega^- \gg 1$ is still valid but $|D^{\pm}| \gg | R^{\pm}|$ in \Eq{Eq:FIDupdown},
and the slow oscillations dominate for
$1/\omega^+ \lesssim t \lesssim 1/\omega^-$. However, the slow oscillations give precisely the same form
as \Eq{Eq:FIDWeights}.
In all cases,  we can  estimate a total \ttwo\ using
$T_2^{-2} = \sum_{n=1}^{n=N} (T_2^{(n)})^{-2}$, where $N \simeq 10^4$ for natural silicon.
However, including both fast ($\omega^+$) and slow ($\omega^-$) terms in the power series, the contributions cancel and the 2-cluster result simply gives a $t^4$ dependence (not observed in experiment) at leading order. Separation of the $\omega^\pm$ timescales is useful not only here, but also potentially in the unmixed ESR regimes of other spin systems.
Further details of the different frequency components of the spin pair-correlation
decays are given in \App{App:pair}.
The analysis for the Hahn case is less straightforward, but nevertheless for (ii),
 we estimate that near NMR-type transitions and 
OWPs, $T_2(\text{Hahn}) \approx 2\times T_2(\text{FID})$ in \App{App:HahnFID}
while  $T_2(\text{Hahn}) \approx  T_2(\text{FID})$ elsewhere.

\begin{figure}[t]
\includegraphics[width=3.5in]{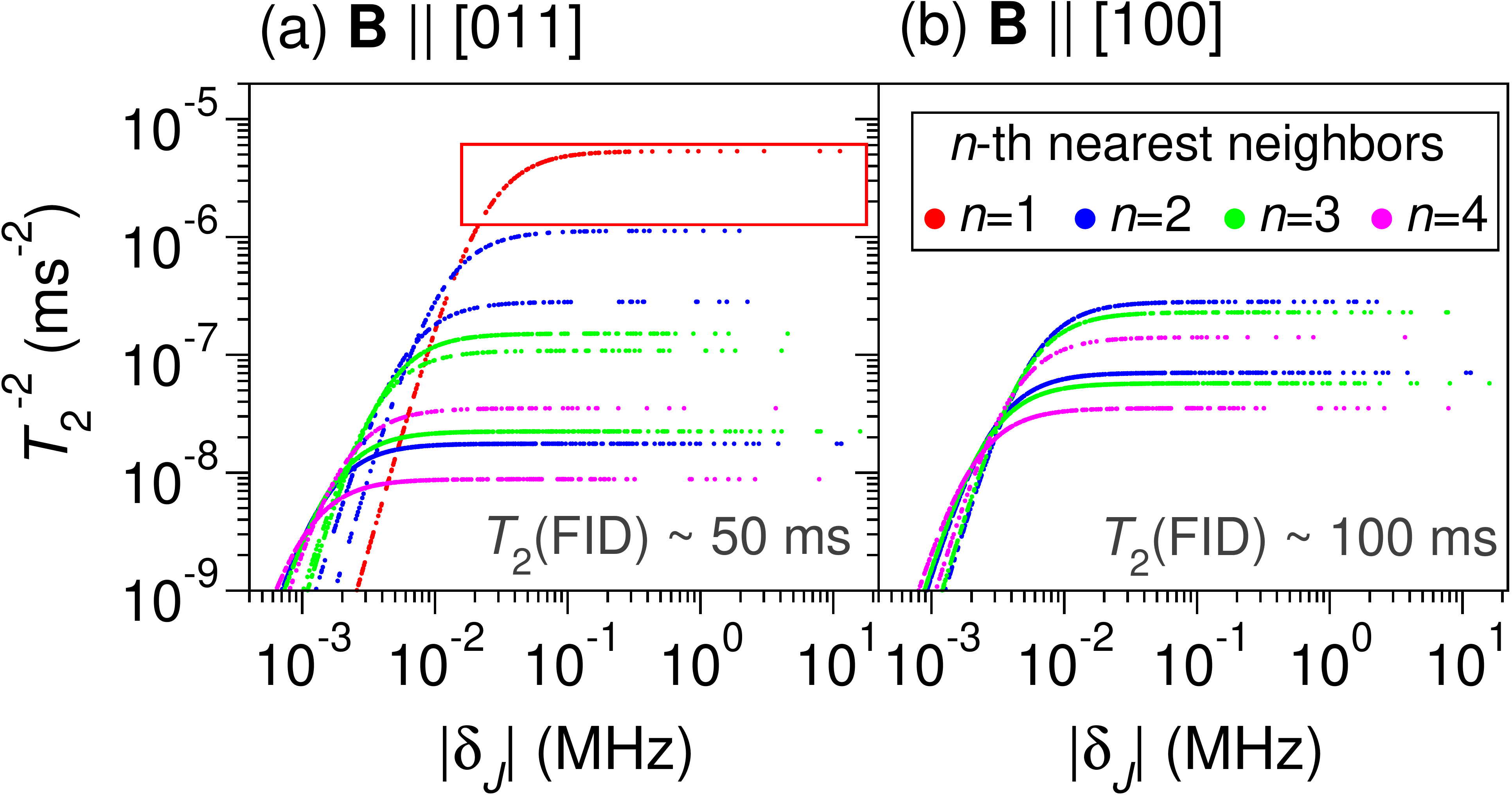}
\caption{(color online) The individual contribution of each spin pair in the bath to the total $(1/T_2)^2$ near OWPs, from \Eq{Eq:FIDWeights}. Data are shown for two magnetic field orientations.
For large $|\delta_J|$, decoherence times become nearly independent of  $|\delta_J|$. The scale of $T_2$ is set by a comparatively small $N \sim 10^2$ set of strongly-coupled spins ($|P_i\delta_J|\gg |C_{12}|$), illustrated in the red box.
$B=79.8$~mT (about $0.1$~mT offset from the OWP) and $P_i\simeq 0.05$. $\gamma_N = 8.465$~MHz/T for \sitwonine.}
\label{Fig:WeightsShort}
\end{figure}

\subsection{Strong coupling approximation}

In \Fig{Fig:WeightsShort}, we use \Eq{Eq:FIDWeights} to evaluate the strength of each $^{29}\text{Si}$ spin pair's individual 
contributions to decoherence of a $^{209}$Bi donor spin in silicon. We plot $1/(T_2)^2$ for each cluster, as a
 function of $|\delta_J|$, in regime (ii) i.e.\ close to OWPs and NMR-type transitions. Strikingly, the spins are grouped into lines of constant $C_{12}$, corresponding to $n$-th nearest neighbor spins.
 Furthermore, for the spin pairs most active in driving decoherence, $1/(T_2)^2$ is only very weakly dependent on $|\delta_J|$.
 The origin of this behavior is clear from \Eq{Eq:FIDWeights}: for large $|P_i\delta_J|\gg |C_{12}|$, the term $|\sin \theta_u -\sin \theta_l| \propto |\delta_J^{-1}|$  while $\omega^+ \propto |\delta_J|$, eliminating
 the dependence on the hyperfine coupling between the central spin and bath spins.

 The insensitivity of the decoherence to the coupling between the central spin and the bath 
might at first seem counter-intuitive. However, the physical origin of this effect is
thus:  increasing the
hyperfine detuning $ \propto |\delta_J^{-1}|$ damps the flip-flopping amplitudes;
 however within this model, the decrease in amplitude
is exactly compensated by a corresponding increase in flip-flop frequency.
 We note that without separation of timescales, 
the $\exp [- t^4]$ decay constants which prevail at times $t \ll \omega_i$ are  dependent
 on $\delta_J^2$.  \cite{Yao2007}
In contrast, our model predicts that a comparatively small number of strongly coupled spins will dominate the 
decoherence, and that their individual contributions to $1/T_2^2$ are approximately equal, although the individual coupling
strengths $ |\delta_J^{-1}|$ vary by orders of magnitude, ranging from $\sim 0.01$ to $10$ MHz. 

To test the validity of this result at $t \sim T_2$ timescales, we run numerical CCE calculations for various field orientations.  
The dipolar coupling, $C_{12}$ is a function of the orientation $\theta$ of the magnetic field and hence the \ttwo\ values vary accordingly.
 For $B \parallel \left\langle 011 \right\rangle$, for example, the $N \sim 10^2$ strongest coupled spin pairs suffice to
 set the scale of \ttwo. We have tested our model by running a 2-cluster CCE calculation with just 120 nearest-neighbor (NN)
spin pairs (e.g.\ for $B \parallel \left\langle 011 \right\rangle$, $C_{12}^{\text{NN}}=1.2$~kHz) which satisfy $|P_i\delta_J|\gg |C_{12}|$, and confirming the calculated \ttwo\ is approximately equal to that considering all $10^4$ spin pairs. For $|P_u| \simeq |P_l|$, we obtain our final expression:
\begin{equation}
T^{u \to l}_2(B,\theta)  \simeq  \overline{C}(\theta) \frac{|P_u(B)|+|P_l(B)|}{\left|P_u(B)-P_l(B)\right|}. 
\label{Eq:T2Formula}
\end{equation}
For most orientations, $\overline{C}(\theta) \approx 4/(C_{12}^{\text{NN}}\sqrt{N})$. However,
 as the magnetic field orientation approaches $B\parallel \left\langle100\right\rangle$, the contribution of nearest-neighbor \sitwonine~spin pairs vanishes, while 2nd- and 3rd-nearest neighbors contribute similarly. Further details of the orientation dependence of \ttwo\ are given in \App{App:Rotation}.

Approaching the high magnetic field limit, ESR-type transitions occur between states where $P_u \simeq - P_l$, such that $T_2\simeq\overline{C}(\theta)$, while for NMR-type transitions as well as OWPs, $P_u \simeq P_l$, and decoherence by the nuclear spin bath is suppressed.\cite{NMRT2}
Finally, we consider a third regime (iii) where one of the $P_i$ is zero, and hence the assumptions made to obtain \Eq{Eq:T2Formula} are not valid. Nevertheless, starting from \Eq{Eq:FIDupdown} we obtain $T_2 \sim \overline{C}(\theta)$ in this regime,
and hence \Eq{Eq:T2Formula} remains a reasonable approximation here.

\begin{figure}[t]
\includegraphics[width=3.5in]{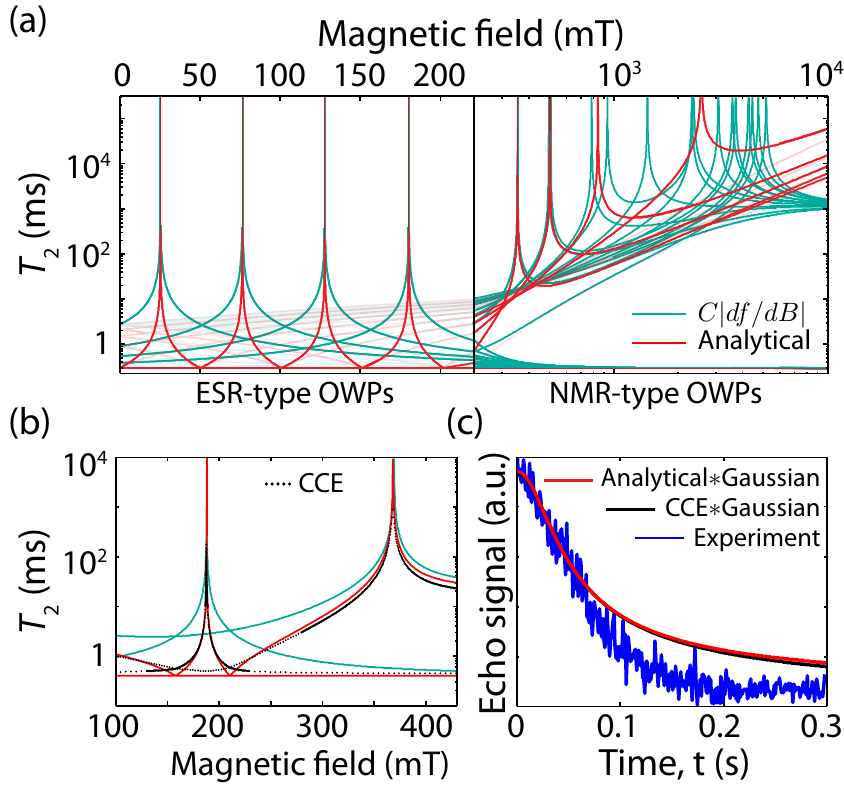}
\caption{(color online) (a) The predicted \ttwo\ values as a function of magnetic field for a variety of allowed transitions in Si:Bi, using \Eq{Eq:T2Formula} derived in the text (labeled `analytical'), show eight OWPs where decoherence is suppressed. We also plot the magnetic field-frequency gradient ($df/dB$); though scaled by an arbitrary constant in order to match the range of estimated \ttwo\ values, the discrepancies with \Eq{Eq:T2Formula} are evident. In the left panel, transitions with no OWP are shown only faintly.
(b) The analytical expression \Eq{Eq:T2Formula} derived in the text is in good quantitative
agreement with CCE numerics, but $df/dB$ is not.
(c) Calculations convolved with Gaussian $B$-field distribution of width 0.42~mT (arising from inhomogeneous broadening from the nuclear spin bath) show an excellent fit with the experimental Hahn echo decay around an ESR-type OWP ($B \sim 80$~mT),\cite{Wolfowicz2013} with no free fit parameters.
}
\label{Fig:OWPs}
\end{figure}

\section{Comparison with experiments and numerics}\label{Sec:Bismuth}

In this section we compare the key result of the paper. \Beq{Eq:T2Formula} is compared with
numerical calculations (including the effect of non-secular terms in $\hat{H}_{\text{int}}$)
as well as new experimental data obtained using bismuth-doped silicon (Si:Bi) as the central spin system.

\subsection{Si:Bi central spin Hamiltonian}

While \Eq{Eq:T2Formula} is in principle valid for a wide variety of 
spin qubits (and encompasses even the limit of unmixed spins) we focus here
on silicon donor qubit systems. In this case, the central spin Hamiltonian is given by:
\begin{equation}
\hat{H}_{\text{CS}}= B\gamma_e \left( \hat{S}_z  + \delta  \hat{I}_z \right) +
A{\bf I} \cdot {\bf S},
\label{Eq:SiBi}
\end{equation}
where $\gamma_e$ is the gyromagnetic ratio of the electron (\mbox{$\gamma_e = 28$}~GHz/T in silicon).
In the particular case of Si:Bi, the donor has 
electron spin $S=1/2$ with isotropic hyperfine coupling $A=1475.4$~MHz to the $^{209}$Bi nuclear spin $I=9/2$,
while $\delta = - 2.488 \times 10^{-4}$ denotes the ratio of nuclear and electronic gyromagnetic ratios.\cite{Mohammady2010}
The energy spectrum of Si:Bi is given in \App{App:SiBiLevels}.
This $\hat{H}_{\text{CS}}$ also applies to other Si donor qubit systems (P, As, Sb)
and results in $2P_i = \Omega_{m} \left(\Omega_{m}^{2} + (I+\tfrac{1}{2})^2 -
m^{2}\right)^{-1/2}$, where
$\Omega_{m}= m + \frac{\gamma_e B}{A}\left(1 + \delta \right)$ and $m=m_S+m_I$ 
is an integer $-|I+S|\leq m \leq I+S$.
Thus, the important mixing parameters $P_i$ in the $T_2$ formula
(\Eq{Eq:T2Formula}) may be evaluated analytically for an arbitrary donor species,
for all field values.

\subsection{Optimal working points}

OWPs in spin donor systems are particular field values
where the $T_2$ times are greatly enhanced.\cite{Mohammady2010,Mohammady2012,Balian2012,Wolfowicz2013}
Here, we use the sensitivity of $T_2$ on magnetic field in the vicinity of OWPs\cite{Balian2012}
as a test of \Eq{Eq:T2Formula}. It is also interesting to investigate deviations from the $T_2 \sim df/dB$ dependence 
that one might expect from classical noise models. 

In \Fig{Fig:OWPs} we plot \Eq{Eq:T2Formula} for Si:Bi for allowed ESR and NMR transitions across a range of magnetic fields.
It shows close agreement with numerical CCE calculations including the effect of $\hat{H}_{\text{int}}$ on $P_i$.
Both \Eq{Eq:T2Formula} and CCE have distinctly different signatures from a curve proportional to $df/dB$, which would be expected in
the case of classical field noise; and they cannot be fitted (except locally) by powers of $df/dB$.

\Bfig{Fig:OWPs}(a) illustrates eight OWPs where $T_2\rightarrow\infty$: four ESR-type and four-NMR type
transitions (these OWPs are all doublets, so there are in fact 16 separate OWP transitions).
The form of \Eq{Eq:T2Formula} clarifies the origin of these discrepancies. For low fields, ($ B \lesssim 1$~T)
the denominator of \Eq{Eq:T2Formula} is $|P_u - P_l| \approx  df/dB$. Thus, it is the numerator ($|P_u| + |P_l|$), 
which accounts largely for the deviation from the form expected for analogous classical noise ($T_2 \propto df/dB$).
However, at higher fields (left panel of \Bfig{Fig:OWPs}(a)), we 
see that while some of the OWPs are coincident with clock transitions where $df/dB \rightarrow 0$, 
others (in particular the NMR-type OWPs) are not. 
 The reason for this deviation is that $\hat{H}_{\text{int}}$ differs from 
 a magnetic field-type term ($\propto (S_z + \delta I_z)$). 
 In other words, while $\hat{H}_{\text{int}}$ determines the form of the interaction between the central spin
and the bath, it is $\hat{H}_{\text{CS}}$ which determines $df/dB$.
 If $\hat{H}_{\text{int}}$ and $\hat{H}_{\text{CS}}$
are of different form, then clock transitions are not OWPs.
In the case of nuclear spin diffusion for Si:Bi systems, for $ B \sim 1$ T, there is
still sufficient mixing between the electronic and nuclear degrees of freedom so that
it is the contact hyperfine interaction ($\propto S_z$) which dominates the effect of $\hat{H}_{\text{int}}$,
thus we may neglect the interaction between the bismuth nuclear spin and the bath,
even for NMR-type transitions. However, in this range, the nuclear Zeeman term 
contributes significantly to $df/dB$ for NMR-type transitions.

In summary, in \Eq{Eq:T2Formula}, it is the denominator ($|P_u - P_l|$) which sets the position of the OWPs:
at these points the bath evolution becomes independent of the state ($\ket{u}$ or $\ket{l}$) of the 
central spin, and so the system-bath entanglement is zero (\Fig{Fig:Pseudospins}). However it is the 
numerator (which can vary by an order of magnitude in the range $0 \leq B \leq 1$~T) which 
provides the most distinct signature of the ``back-action'' between quantum bath and central spin.

\subsection{Comparison with experiment}

The donor ESR line is inhomogeneously broadened by unresolved coupling to \sitwonine, leading to an effective
Gaussian magnetic field variation across the ensemble (FWHM of $0.42$~mT for Bi in natural silicon).
Therefore, to predict the measured \ttwo\ at an ESR-type OWP we convolve \Eq{Eq:T2Formula} with
 the corresponding Gaussian magnetic field profile (this also takes care of the divergence in
 \ttwo\ at the OWP). This is found to give a non-Gaussian decay and reaches its $e^{-1}$ value
 at $100$~ms as shown in \Fig{Fig:OWPs}(c) in close agreement with the experimental value of 93 ms.\cite{Wolfowicz2013}
Details of the convolution are given in \App{App:Convolution}.

\begin{figure}[t]
\includegraphics[width=3.5in]{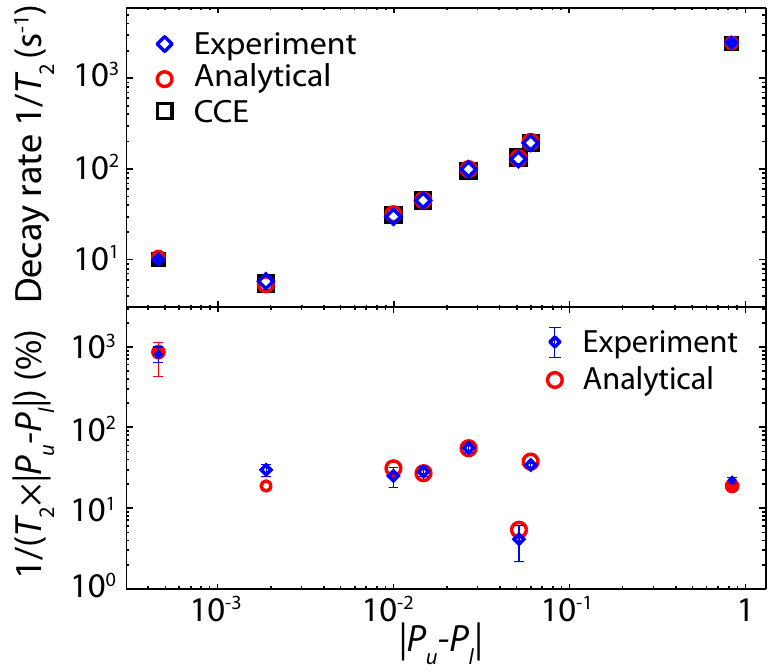}
\caption{(color online) Comparison between theoretically predicted and measured
 \ttwo\ in \textsuperscript{nat}Si:Bi for various transitions,
showing remarkable agreement across a wide range of mixing regimes $|P_u-P_l|$. The label `analytical' refers to \Eq{Eq:T2Formula}. Measurements were made at 4.8~K using ESR with a microwave frequency of 9.77 or 7.03~GHz (filled symbols), or electron-nuclear double resonance (ENDOR) between 200~MHz and 1~GHz using the method described in Ref.~\onlinecite{Morton2008} (empty symbols), at magnetic fields between 100 and 450~mT. These parameters are all in the regime where $|P_u-P_l|\approx df/dB$. The Bi donor concentration was $\leq 10^{16}$~cm$^{-3}$, and decoherence times are limited by \sitwonine\ spin diffusion. The theoretical points are based on a predicted value for $\overline{C}(\theta) = 0.42$~ms. In the lower panel, the decay rates are normalized by $|P_u-P_l|$ to highlight the effect of $|P_u|+|P_l|$, and shown relative to the case
when $|P_u|=|P_l|$. Further experimental details are in \App{App:Rotation}.}
\label{Fig:ExperimentTheory}
\end{figure}

We move on to test \Eq{Eq:T2Formula} across a broader range of parameters (\Fig{Fig:ExperimentTheory}),
 by comparison with \ttwo\ measurements of ESR transitions towards the high-field regime ($|P_u-P_l|\simeq1$),
 and \ttwo\ for a variety of different NMR transitions where $|P_u-P_l|$ varies by two orders of magnitude. Both CCE and \Eq{Eq:T2Formula} give excellent agreement with the measured values.
 The primary variation in \ttwo\ is due to the $|P_u-P_l|$ term; this is divided out in the lower panel of \Fig{Fig:ExperimentTheory},
 where the additional variations due to $|P_u|+|P_l|$ are apparent in the experiment.

We emphasize that the derivation of \Eq{Eq:T2Formula} involves a range of approximations.
Assumptions have been made regarding the strong coupling approximations and the importance of certain spins.
Only pair-correlations are considered which means that numerics are not converged for 
Hahn echo decays near OWPs. Thus, while one might expect a factor of two agreements with experimental 
comparisons, the agreement with the data over such a large range is remarkable and indicates that
the form of $T_2$ predicted by \Eq{Eq:T2Formula} will persist even for higher-order CCE calculations.

\section{Conclusions}\label{Sec:Conclusion}

In summary, we have shown that a field dependence given by $T_2(B) \propto \left(|P_u|+|P_l|\right) \left(|P_u-P_l|\right)^{-1}$, distinctly different from classical field noise which yields $T_2(B) \propto df/dB$,\cite{Ithier2005,Martinis2003,Mohammady2012}
is a generic and robust feature of mixed electron-nuclear spin systems, valid over a broad range of ESR and NMR transitions both close to and far from OWPs. The range also includes the {\em unmixed} case in the limit $|P_u| = -|P_l|$.

In addition to use of an OWP, decoherence by nuclear spin diffusion can be suppressed by enrichment of the host using a spin-zero isotope (e.g.\ using enriched \sitwoeight)\cite{Tyryshkin2012}. The effect of reducing the nuclear spin concentration on \ttwo\ is explicit in the $\overline{C}(\theta)$ term, but it also causes narrowing of the ESR linewidth and hence reduces the effective magnetic field distribution to a narrower range around the OWP. As the nuclear spin concentration becomes negligible, other decoherence processes become dominant, including couplings to other (e.g.\ donor) spins which can similarly be analyzed for a quantum-correlated bath.\cite{InPreparation}
It has been previously demonstrated for P donors in Si that line-broadening effects caused by nuclear spins in the bath nuclei suppress donor-donor flip-flops,\cite{Witzel2010} thus future studies must consider partial isotopic enrichment and a mixture of donor-donor and $^{29}\text{Si}$-related decoherence mechanisms.

\begin{acknowledgements}

We gratefully acknowledge very helpful advice from Ren Bao Liu and Wayne Witzel. We also acknowledge fruitful discussions with Hamed Mohammady, Gavin Morley, Chris Kay and Alexei Tyryshkin. This research is supported by the EPSRC through the Materials World Network (EP/I035536/1) and a DTA, as well as by the ERC under FP7/2007-2013 / ERC grant agreement no. 279781. JJLM is supported by the Royal Society.

\end{acknowledgements}

\appendix

\section{Analysis of spin pair-correlations}\label{App:pair}
 
Here we consider the contributions which dominate the spin pair-correlation
in different regimes and timescales.
We begin with our FID for the $n$-th spin pair, as given in the main text:
\begin{eqnarray}
\mathcal{L}^{u\to l}_{n,\text{FID}}(t) &=&
D^+e^{-i\omega^- t} + D^-e^{+i\omega^- t} \nonumber \\
           && ~~+ R^+e^{-i\omega^+t} + R^-e^{+i\omega^+t},
\label{Eq:FID1}
\end{eqnarray}
where \mbox{$ R^{\pm} = \frac{1}{2}\sin \theta^-(\sin \theta^- \mp \sin \theta^+)$} and
\mbox{$ D^{\pm} =  \frac{1}{2}\cos \theta^-(\cos \theta^- \pm  \cos \theta^+)$}
and with $\theta^\pm= \frac{1}{2} (\theta_{u}\pm \theta_{l})$ and
$\omega^{\pm}=\omega_u \pm \omega_l$.
Noting that $\omega^+ \gg \omega^-$ we infer that the $R^\pm$ terms act on very
different timescales from the terms proportional to $D^\pm$. We consider the $R^\pm$ and $D^\pm$ terms
separately.

For either of the thermal $\ket{k_\mathcal{B}}=\sket{\uparrow\downarrow}$ or $\sket{\downarrow\uparrow}$ bath states, 
if we set $\omega^-=0$, we obtain the fast oscillating contribution:
\begin{eqnarray}
|\mathcal{L}_{n,\text{FID}}^{u \to l}(\ket{k_\mathcal{B}},t)|^2 &\simeq& 1-4\left(D^+ + D^-\right)\left(R^+ + R^-\right) \sin^2 \tfrac{\omega^+t}{2} \nonumber \\
 &&{}- 4R^+ R^-  \sin^2 (\omega^+ t).
\label{Eq:fast}
\end{eqnarray}

We extract the contribution of each cluster to the
total decoherence by means of a power-expansion; for short times we obtain
$|\mathcal{L}_{n}^{u \to l}(\ket{k_\mathcal{B}},t)|^2 \approx 1- a_2t^2 \approx 1-(2t/T^{(n)}_2)^2 \exp{\left[-(2t/T^{(n)}_2)^2\right]}$, yielding the $n$-th cluster contribution to $T_2$:
\begin{equation}
\left(T^{(n)}_{2}\right)^{-2}  \approx \left[\left(D^+ + D^-\right)\left(R^+ + R^-\right)+ 4R^+ R^-\right]
\left(\omega^+\right)^2.
\label{Eq:T2Ka}    
\end{equation}

If we then make the strong coupling approximation, and average over bath states, as in the main text,
the weights in \Eq{Eq:T2Ka} can also be written as:
\begin{equation}
\frac{1}{(T_2^{(n)})^2}  \simeq \frac{(\theta_u - \theta_l)^2}{4^2} (\omega^+)^2.
\label{Eq:T2K}
\end{equation}

Then, noting $\theta_i \approx C_{12}/\omega_i$ and $\omega^+ \approx \delta_J (|P_u|+|P_l|)$
we easily obtain $\frac{1}{T_2^{(n)}} \propto  \frac{|P_u- P_l|}{|P_u|-|P_l|}$,
for the cases $ |P_u| \simeq |P_l|$, which include both 
the unmixed ESR limit as well as the NMR
and OWP limits.

\begin{figure}[t!]
        \includegraphics[width=3.5in]{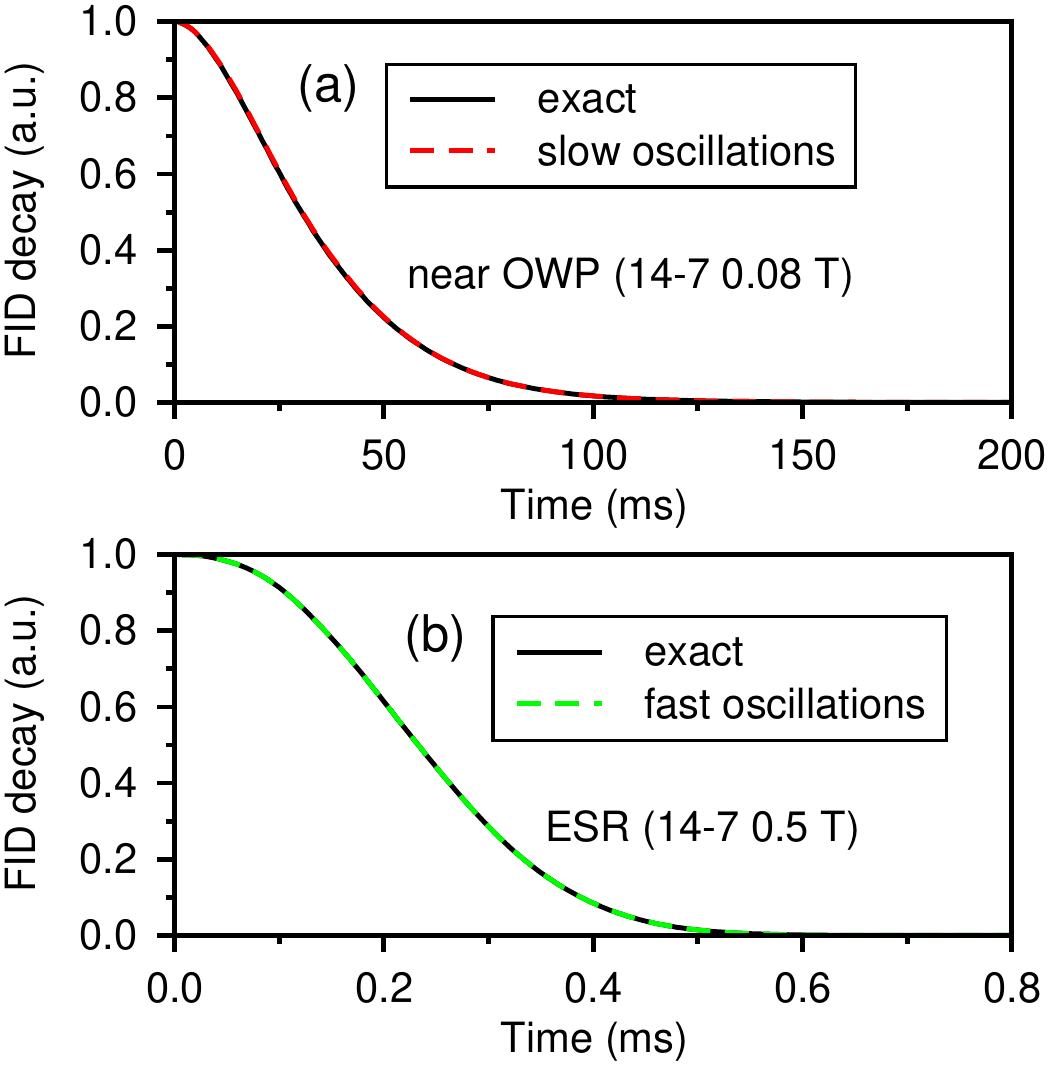}
    \caption{Shows that OWP regimes are dominated by slow oscillating terms while ESR regimes
are dominated by fast oscillating terms in \Eq{Eq:FID1}.
(a) Compares decays obtained from \Eq{Eq:FID1} (exact) with decays obtained from \Eq{Eq:slow}
(slow oscillations only). (b) Compares decays obtained from \Eq{Eq:FID1} (exact) with decays obtained from \Eq{Eq:fast}
(fast oscillations only)}
    \label{Fig:Timescales}
\end{figure}

 However, care is needed when considering OWP and NMR regimes since here,
 $P_u \simeq P_l$, $\theta_u \simeq \theta_l$
and thus $D^\pm \gg R^\pm$. Here, $D^++D^- \to 1$ while $R^\pm \to 0$.
 Decay timescales become long and comparable to $1/\omega^-$ while the $R^\pm$ amplitudes are negligible
and thus the slow oscillating components are important. In that case, we would, in contrast to \Eq{Eq:fast},
neglect the fast oscillations. Then we obtain,
\begin{eqnarray}
|\mathcal{L}_{n,\text{FID}}(\ket{k_\mathcal{B}},t)|^2 \simeq 1-  4D^+ D^-  \sin^2 \omega^- t.
\label{Eq:slow}
\end{eqnarray}
In this case, $(T_2^{(n)})^{-2}  \approx D^+ D^- \left(\omega^-\right)^2$. 
However, since
 \begin{eqnarray}
\left[\left(D^++ D^-\right)\left(R^++ R^-\right)+ 4R^+ R^-\right]
\left(\omega^+\right)^2 \nonumber \\
\to  D^+ D^- \left(\omega^-\right)^2
\label{Eq:equiv}
\end{eqnarray}
as  $P_u \to P_l$, the contribution to $1/T_2^2$ from each cluster, in fact, still has the same form
as \Eq{Eq:T2Ka}. In other words, the relative weights obtained from the slow, high-amplitude
contributions are quite similar to those obtained by considering the faster, lower oscillations
and thus the $T_2$ expression \Eq{Eq:T2Formula} is still valid.
 
We recall that if we attempt to estimate $1/T_2^{(n)}$ from the short-time behaviour of the exact expression,
without separating timescales, to leading order, an $\exp{[t^{-4}}]$ decay is obtained, rather than the 
observed $|\mathcal{L}| \approx e^{-{(t/T_2)^2}}$. Thus it is necessary to
consider the slow and fast oscillating terms {\em separately} if estimating analytical values
of $T_2$ from the calculated decay functions.

\Bfig{Fig:Timescales} clarifies this. Here we show the full temporal decay for all pairs
\begin{equation}
\mathcal{L}^{u \to l}(\ket{k_\mathcal{B}},t) = \prod_{n} \mathcal{L}^{u \to l}_{n}(\ket{k_\mathcal{B}},t),
\label{CCE}
\end{equation}
where $\mathcal{L}_n^{u \to l}(\ket{k_\mathcal{B}},t)$ is given by \Eq{Eq:FID1}
and compare with (a) the slow terms in an OWP regime (\Fig{Fig:Timescales}(a)) where
$\mathcal{L}_n^{u \to l}(\ket{k_\mathcal{B}},t)$ is given by \Eq{Eq:slow}
and (b) the fast terms in the ESR regime (\Fig{Fig:Timescales}(b))
where $\mathcal{L}_n^{u \to l}(\ket{k_\mathcal{B}},t)$ is given by \Eq{Eq:fast}.

\Bfig{Fig:Timescales} shows that while the fast terms completely dominate coherence decay in the ESR regime,
the slow terms completely dominate the decays in the OWP/NMR regime yet the form of the weights in the
power expansion is similar: if added, the two contributions thus cancel (albeit briefly) yielding the quartic decay.
This $\exp{[-t^4]}$ decay is of course valid on extremely short timescales $t \ll (\omega^+)^{-1}$ but
not on the $T_2$ timescale.

Thus, when inferring decay rates on the $T_2$ timescale from the early time behavior,
it is important to consider different frequency components separately.
In fact, even the fast oscillation behavior is not entirely straightforward.
For the slow oscillations, \Eq{Eq:slow} involves a single 
frequency and an 
approximate $\exp{[-(t/T_2)^2]}$ decay  
is straightforwardly  inferred.

For the fast oscillations however, \Eq{Eq:fast} may be rewritten as follows:
\begin{eqnarray}
|\mathcal{L}_{n,\text{FID}}(\ket{k_\mathcal{B}},t)|^2& \simeq & 1- \sin^2 \theta^- \cos^2 \theta^+ \sin^2(\omega^+ t)  \nonumber \\
 &-&  \sin^2 2\theta^- \sin^2\tfrac{\omega^+t}{2}\nonumber \\
 &-& \frac{1}{4}\sin^2 2\theta^- \sin^2 (\omega^+ t)  \nonumber \\
&=& 1-L_s(t)-\left(L_{l1}(t)-L_{l2}(t)\right).\nonumber \\
\label{Eq:fast1}
\end{eqnarray}
We see that it combines three separate interfering terms, where $L_{l1}$ oscillates at half the frequency of 
the others. In fact, a power expansion of either one of the individual terms $L_s(t)$, $L_{l1}(t)$ and $L_{l2}(t)$
would yield the same 
$(T_2^{(n)})^{-2} \simeq \frac{1}{4^2}\left( \theta_u -\theta_l\right)^2 \left(\omega^+\right)^2$
which leads to our $T_2$ expression. It is the ubiquitous nature of this 
$ \left( \theta_u -\theta_l\right)^2 \left(\omega^+\right)^2$ term which underlies the robustness of the
experimentally observed $T_2 \sim \frac{|P_u| +|P_l|}{|P_u-P_l|}$ behavior.

We note that it is in fact the term
$L_s(t)= \frac{1}{4}\left(\sin \theta_u -\sin \theta_l\right)^2 (\omega^+)^2$
 which yields a quadratic dependence at short times.
However, numerics show that it is the
 $1 -(L_{l1}(t)+L_{l2}(t))$ terms  which 
overwhelmingly determine the decay on longer
 $T_2$ timescales (but actually make little contribution on the $t \ll (\omega^+)^{-1}$ timescale,
where there is once again a brief cancellation of these
near equal amplitude oscillations).

The Landau-Zener (LZ) regimes (there are four such regions for Si:Bi) do not fit the above analysis,
which assumed $|P_u| \simeq |P_l|$. For the LZ points either $P_u \simeq 0$ or $P_l \simeq 0$.
Thus, assuming $P_u \simeq 0$ we obtain, $|\mathcal{L}^{u \to l}_{n,\text{FID}}(t)|^2 \simeq 1- \sin^2 \theta_u \sin^2 \omega_u t$
and hence for $ t \ll (\omega_u)^{-1} $, we have simply $|\mathcal{L}^{u \to l}_{n,\text{FID}}(t)|^2 \simeq  1- C_{12}^2t^2$.
Hence, we obtain $T_2 \simeq C(\theta)$ as in the main text after the usual bath average and sum over clusters.

\section{Relation between Hahn echo and FID}\label{App:HahnFID}

\begin{figure}[t!]
\includegraphics[width=3.5in]{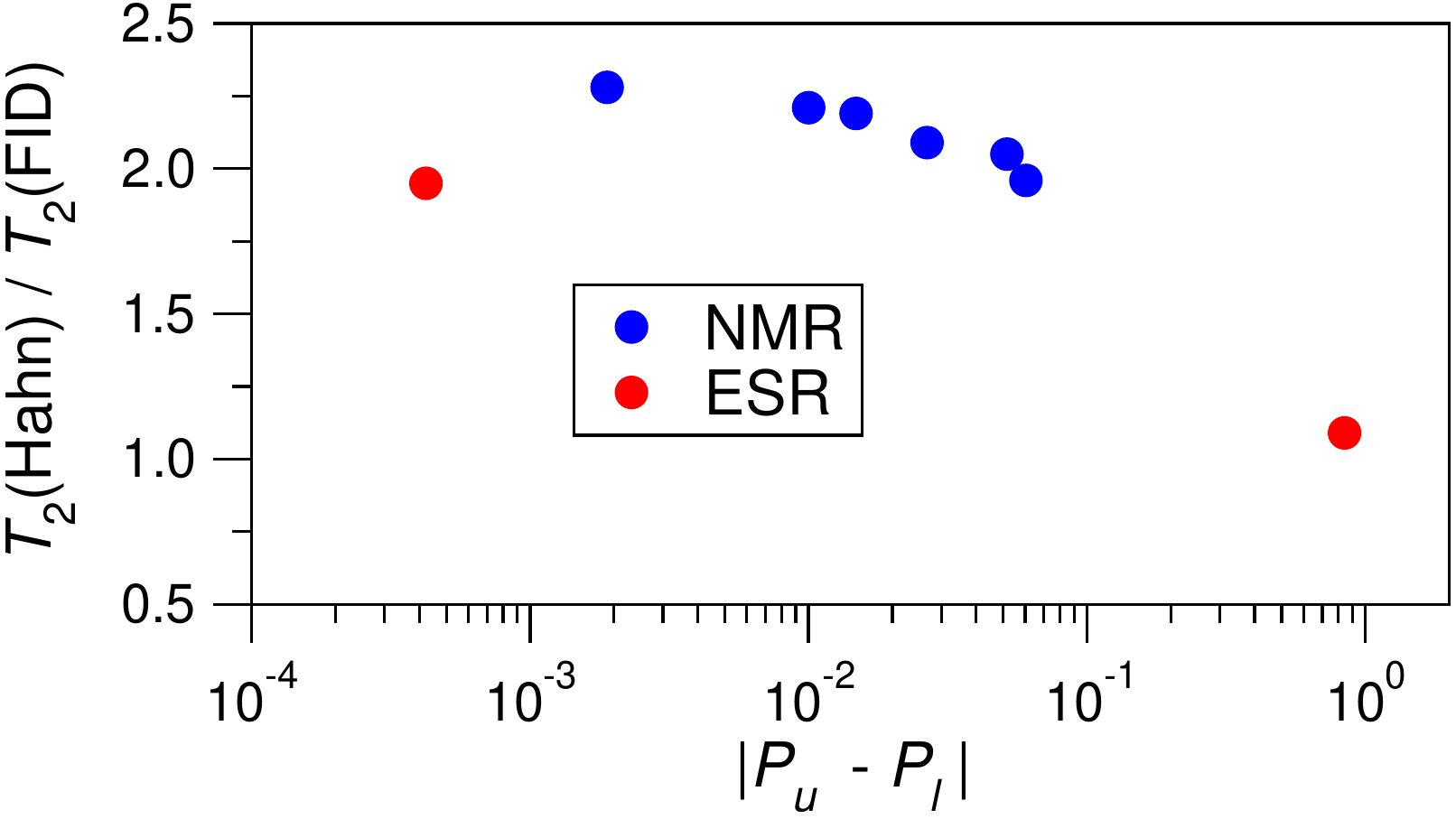}
\caption{(color online) Comparison of calculated $T_{2}(\text{Hahn})$ and $T_{2}(\text{FID})$
for the various ESR and NMR-type transitions of Si:Bi in \Fig{Fig:ExperimentTheory}.
Near OWPs (where $\left|P_u - P_l\right| \ll 1$), $T_{2}(\text{Hahn}) / T_{2}(\text{FID}) \simeq 2$.
}
\label{fig:HahnVsFID}
\end{figure}

While FID and Hahn echo decays are generally of the same order, within about $5$mT of 
an OWP, our calculated Hahn echo (pair-correlations) shows non-decaying
oscillatory behavior at timescales beyond a few ms, indicating loss of numerical convergence.  In contrast, the FID exhibits no such problems and shows converged, near-Gaussian decays to zero intensity for all timescales and magnetic
fields. Nevertheless, there is always a period of initial near-Gaussian decay from which we
extract $T_{2}(\text{Hahn})$. This initial period of convergence is extended to
longer times as higher order cluster contributions are taken into
account \cite{Witzel2012}. Based on the above, we estimate numerically the
ratio $T_{2}(\text{Hahn}) / T_{2}(\text{FID})$ and
confirm in \Fig{fig:HahnVsFID} that
$T_{2}(\text{Hahn}) / T_{2}(\text{FID}) \approx 2$
near OWPs (where $\left|P_u - P_l\right| \ll 1$).

\section{Dependence of $T_2$ on crystal orientation}\label{App:Rotation}

The strength of the dipolar interaction $C_{12}$ depends on the angle between the vector joining the interacting spins and the direction of the magnetic field ${\bf B}$. As a result, $T_2$ varies with the orientation of the crystal sample relative to ${\bf B}$ \cite{DeSousa2003_1,Witzel2006,Tyryshkin2006,George2010}.
The dipolar prefactor $\overline{C}(\theta)$ in our analytical $T_2$ formula (\Eq{Eq:T2Formula}) depends on $C_{12}$ and is thus a function of crystal orientation. The prefactor is defined as
\begin{equation}
\overline{C}(\theta) = \frac{4}{\sqrt{\sum_s{ N_s \left(C_{12}^{(s)}\right)^2  }}},
\label{Eq:DipolarPrefactor}
\end{equation}
where $s$ labels a unique value of spin pair dipolar strength $C_{12}^{(s)}$, or ``shell'', which occurs $N_s$ times. We see below that including shells up to $s=3$ gives a good estimate of $\overline{C}(\theta)$, although for most angles $s=1$ suffices. 

\begin{figure}[t!]
\includegraphics[width=3.5in]{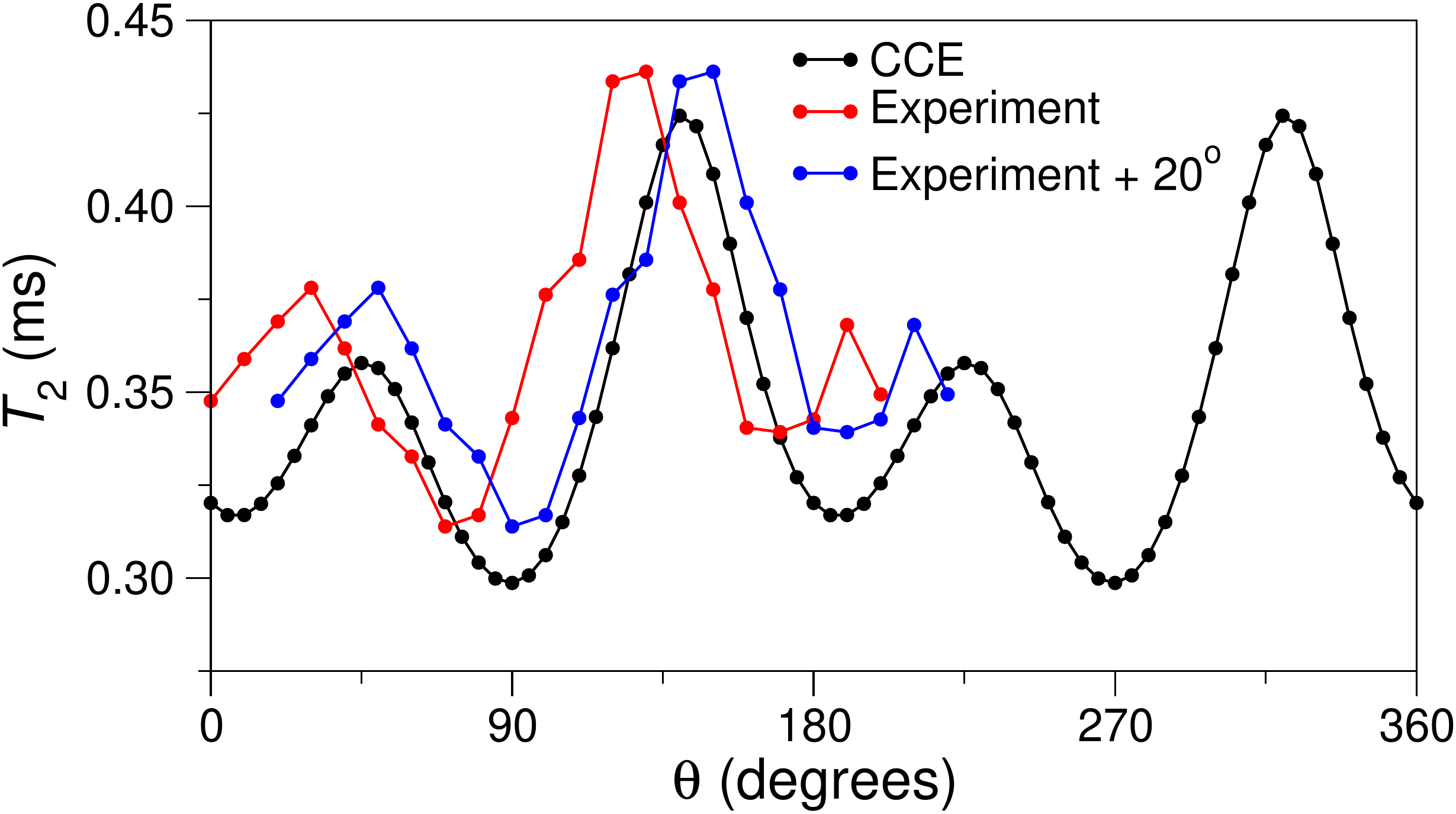}
\caption{(color online) Angular dependence of $T_2$ for an ESR transition of Si:Bi. Rotation was performed about the $[\overline{1} \overline{1} 2]$ axis in the $[\overline{1} 1 0]$ - $[1 1 1]$ plane with $\theta$ from $[\overline{1} 1 0]$. The best match to experiment was obtained for a $5^{\circ}$ tilt in the rotation axis and a zero-offset of $20^{\circ}$.}
\label{Fig:Rotation}
\end{figure}

In order to estimate the value of $\overline{C}(\theta)$ for the experimental data in \Fig{Fig:ExperimentTheory}, $T_2$ was measured (\Fig{Fig:Rotation}) as a function of crystal orientation. X-ray diffraction using the back-reflection Laue technique showed the rotation axis to be close to $[\overline{1} \overline{1} 2]$. The external magnetic field is in the rotation plane, defined by the angle $\theta$ such that $\theta=0^{\circ}$ and $\theta=90^{\circ}$ correspond to the field parallel to $[\overline{1} 1 0]$ and $[1 1 1]$ respectively. The value for $\overline{C}(\theta)$ was determined taking into account uncertainties in both the initial angle $\theta = 0^{\circ}$ and a slight tilt of the rotation axis from $[\overline{1} \overline{1} 2]$. The best match to experiment was obtained for the rotation axis tilted about $[1 1 1]$ by $5^{\circ}$ and a $20^{\circ}$ shift in $\theta$. 

Most of the data points in \Fig{Fig:ExperimentTheory} were measured for $\theta=135^{\circ}$, corresponding to $\overline{C}(135^\circ) = 0.40$~ms, using a sample with [Bi]$=3\times10^{15}$~cm$^{-3}$. The exception is the ESR-type OWP point (with the lowest value for $|P_u-|P_l|$), which was measured with a different sample with the field aligned along $[011]$ and  [Bi]$=10^{16}$~cm$^{-3}$. This gives the same value for $\overline{C}(\theta)$, allowing us to use it for comparison with the NMR points in the main text. The small difference in [Bi] is not expected to affect decoherence times in the regimes studied, which are instead dominated by nuclear spin diffusion.

\begin{figure*}[t!]
\includegraphics[width=7.0in]{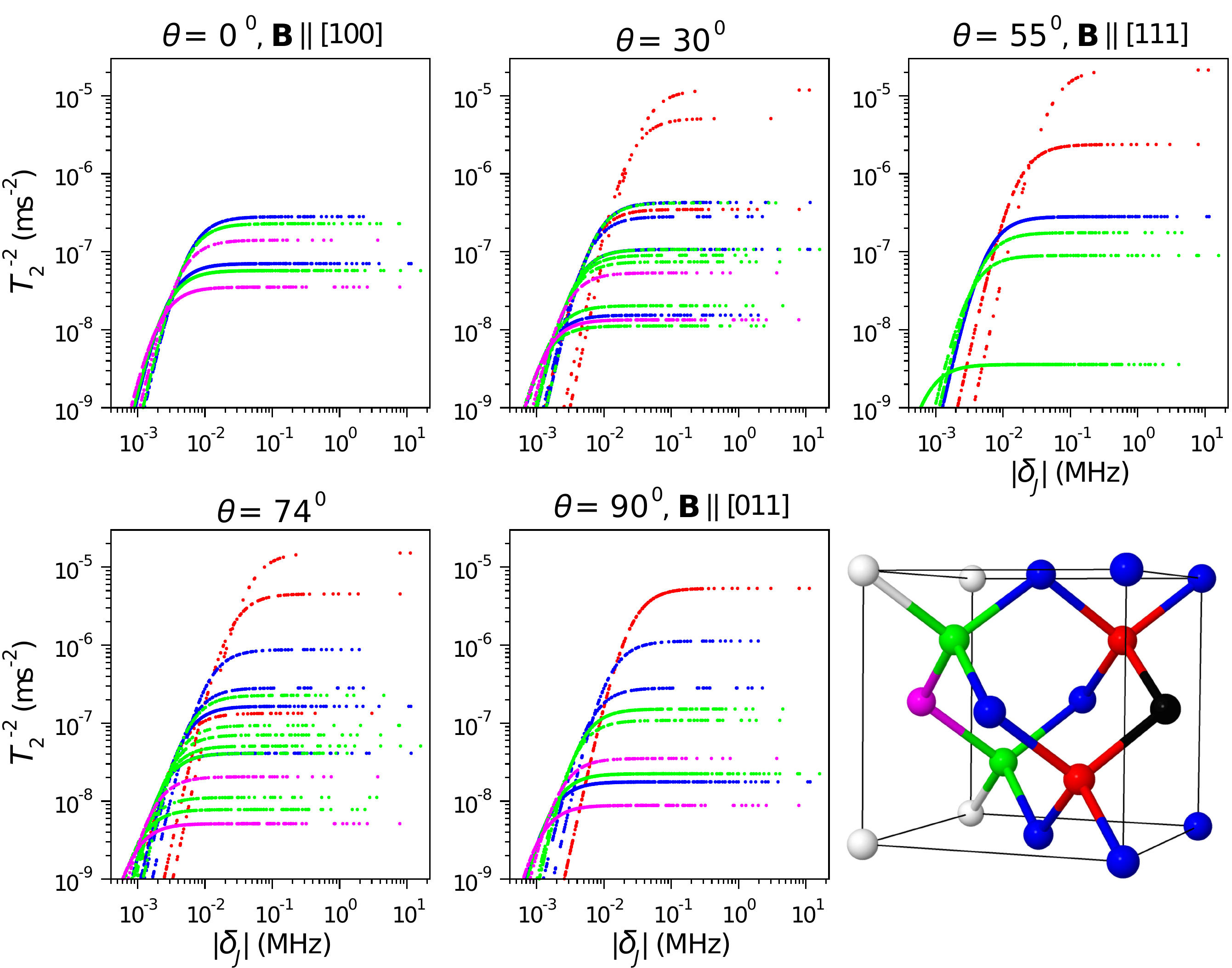}
\caption{(color online) Theoretical contributions of spin pairs to $T_2(\text{Hahn})$, colored according to $n$-th nearest neighbors relative to the black nucleus as illustrated in the last panel.
First nearest neighbors dominate decoherence for
rotation angles $\theta \gtrapprox 30^{\circ} $.
At $\theta=0^{\circ}$, first nearest neighbor contributions are diminished and second and third nearest neighbors contribute the most to $T_2$.
Rotation is performed about $\left[01\bar{1}\right]$ in the $\left[011\right]- \left[100\right]$ plane,
with $\theta$ from $\left[100\right]$.
}
\label{Fig:WeightsLong}
\end{figure*}

We now proceed to determine the full angular dependence of $\overline{C}(\theta)$. The various $1/T_{2}^2$ contributions of \textsuperscript{29}Si spin pairs as a function of crystal rotation angle are shown in \Fig{Fig:WeightsLong}. The data in \Fig{Fig:WeightsLong} was generated from \Eq{Eq:FIDWeights} near the ESR-type OWP (rotation around $\left[01\bar{1}\right]$) of \textsuperscript{nat}Si:Bi, however, our results are independent of $B$ and the central donor species, up to a scaling factor on 1/$T_2^2$ contributions. 

In \Fig{Fig:WeightsLong}, the different shells are labeled according to whether the interacting spins are first,
second, third or fourth nearest neighbors (1--, 2--, 3--, 4--NNs). The total $T_2$ is obtained by summing $1/T_2^2$ contributions from all spin pairs in the bath. We pick the strongest $N$ spin pairs (i.e., those with the largest $1/T_2^2$ contribution) such that the sum over $1/T_2^2$ is about $70 - 80\%$ of the total $T_2$, and find that $N \simeq 270$ for $\theta = 0^{\circ}$ and $N \simeq 100$ for all the other rotations considered. Contributions from 1--NNs are dominant for $\theta \gtrapprox 30^{\circ}$. In \Table{Table:Weights}, we show that 1--NNs suffice to set the scale of $T_2$ for $\theta \gtrapprox 30^{\circ}$ by comparing $\overline{C}(\theta)$ obtained from only 1--NNs to $\overline{C}(\theta)$ extracted from numerical CCE2 $T_2$ and using \Eq{Eq:T2Formula}. For $\theta = 0^{\circ}$, 2--NNs and 3--NNs contribute the most, without any 1--NNs being involved in setting the scale of $T_2$. Including only the strongest 2--NN and 3-NN contributions, for $\theta = 0^{\circ}$ we find $\overline{C}(0^{\circ}) \simeq 0.97$~ms, compared to $\overline{C}(0^{\circ}) = 1.1$~ms obtained using the numerical $T_2$. Thus, using the estimated $\overline{C}(\theta)$ values in the first column of \Table{Table:Weights}
provides a reasonable estimate of the dipolar prefactor $\overline{C}(\theta)$ as a function of crystal rotation.

\begin{center}
\begin{table*}[!t]

\begin{tabular}{c|c|c|} \hline \hline

Rotation angle $\theta$ (degrees) & 1-NN contribution to $\overline{C}(\theta)$ (ms)& Numerical $\overline{C}(\theta)$ (ms)\\ \hline

$90$   &   $0.37$   &  0.40 \\ \hline
$74$ &   $0.35$   &  0.39 \\ \hline
$55$   &   $0.32$   &  0.37 \\ \hline
$30$ &   $0.41$   &  0.45 \\ \hline
$0$ & None. 2--NN, 3--NN: $0.97$ & $1.1$ \\ \hline
    
\end{tabular}

\caption{Numerical values of the dipolar prefactor $\overline{C}(\theta)$ compared to $\overline{C}(\theta)$ when including only nearest neighbor spin pairs,
demonstrating that first nearest neighbors set the scale of $T_2$ for rotation angles $\theta \gtrapprox 30^{\circ}$. For $\theta = 0^{\circ}$, 1--NNs do not contribute at all and 2,3--NNs largely determine $T_2$. The total number of strongest spin pairs for each orientation was chosen such that the $T_2$ obtained was about $70 - 80\%$ of the total $T_2$ when including all spin pairs in the bath. Rotation is performed about $\left[01\bar{1}\right]$ in the $\left[011\right]- \left[100\right]$ plane,
with $\theta$ from $\left[100\right]$.
}
\label{Table:Weights}
\end{table*}
\end{center}

\clearpage

\section{${\bf Si:Bi}$ energy spectrum}\label{App:SiBiLevels}

\begin{figure}[!t]
\includegraphics[width=3.5in]{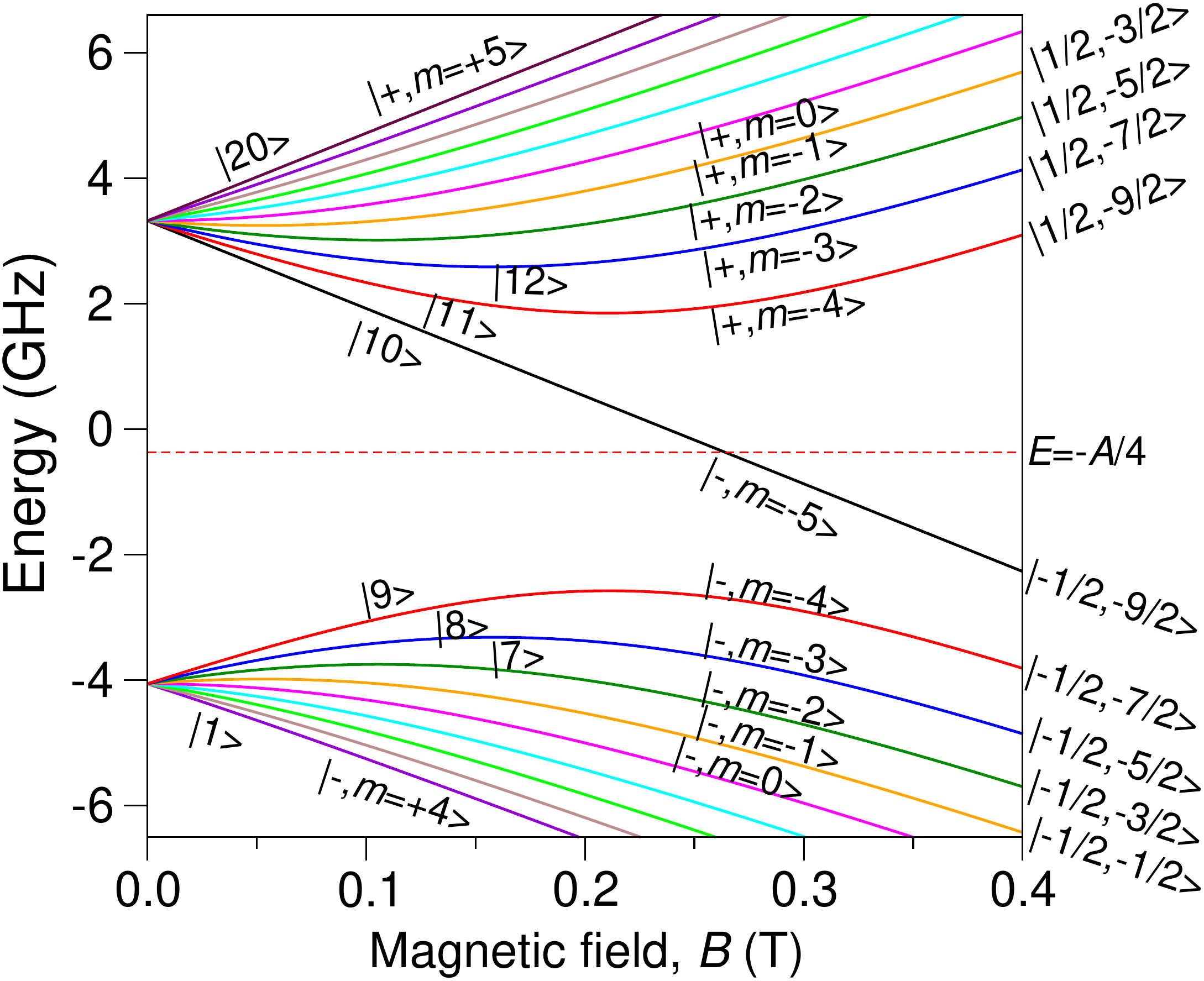}
\caption{(color online) Energy spectrum of Si:Bi. The eigenstates can be labeled in order of increasing energy ($\ket{i}$, $i=1,2,\dots,20$),
in the Zeeman basis
($\ket{m_S,m_I}$, $m_S = \mp\tfrac{1}{2}$, $m_I = -\tfrac{9}{2},-\tfrac{7}{2},\dots,\tfrac{7}{2},\tfrac{9}{2}$),
or the adiabatic basis ($\ket{\pm,m}$, $-5 \leq m \leq 5$).\cite{Mohammady2010}
We refer to $\ket{\pm,m} \leftrightarrow \ket{\mp,m-1}$ and $\ket{\pm,m} \leftrightarrow \ket{\pm,m-1}$ as ESR and NMR-type transitions respectively noting that $\ket{-,m} \leftrightarrow \ket{+,m-1}$ are forbidden at high fields.\cite{Mohammady2010}
$A$ is the strength of the isotropic electron-nuclear hyperfine interaction.
}
\label{Fig:SiBi}
\end{figure}

The Si:Bi spin system is an electron ($S=1/2$) coupled to a nucleus with $I=9/2$ giving a 20-dimensional Hilbert space.\cite{Mohammady2010} The effective Hamiltonian is given in \Eq{Eq:SiBi} and has the energy spectrum shown in \Fig{Fig:SiBi}.

\section{Convolution of $T_2$ formula}\label{App:Convolution}

We note that \Eq{Eq:T2Formula} gives divergent $T_2$ values at the OWP; comparison with CCE indicates that it becomes unreliable within $\sim 0.01$~mT of the OWP and non-secular terms cap the maximum $T_2 \lesssim 10$ seconds. However, for Bi donors in natural silicon, the ESR line is inhomogeneously broadened by unresolved coupling to \textsuperscript{29}Si, leading to an effective Gaussian magnetic field variation across the ensemble with FWHM of $0.42$~mT. Therefore, we are nevertheless able to use \Eq{Eq:T2Formula} to predict the measured $T_2$ at an ESR-type OWP by convolving it with the appropriate magnetic field distribution:
\begin{equation}
D(t) = \frac{1}{w \sqrt{2\pi}}
\int{ e^{\frac{-\left(B-B_{\text{OWP}}\right)^2}{2w^2}} e^{-\left(t/T_2\right)^2} dB}
\end{equation}
where $w=0.21$~mT and $T_2$ has a dependence on $B$, as given by \Eq{Eq:T2Formula}. The convolution $D(t)$ is found to give a non-Gaussian decay, and reaches its $e^{-1}$ value at $100$~ms as shown in \Fig{Fig:OWPs}(c), in close agreement with the experimental value of $93$~ms for Si:Bi OWP 2~\cite{Wolfowicz2013}. The convolution sums $T_2(B)$ contributions which vary over orders of magnitude and thus represents a sensitive test of \Eq{Eq:T2Formula} around an ESR-type OWP.

\bibliography{refs}

\begin{thebibliography}{41}
\bibitem[{NMR()}]{NMRT2}
\bibinfo{note}{{In regimes where we can neglect the effect of the central
  nucleus in $\hat{H}_{\text{int}}$.}}
\end{thebibliography}

\end{document}